# Integrated quantitative PIXE analysis and EDX spectroscopy using a laser-driven particle source

## Laser-driven sources for materials analyses

F. Mirani*[1], A. Maffini[1], F. Casamichiela[1], A. Pazzaglia[1], A. Formenti[1], D. Dellasega[1], V. Russo[1], D. Vavassori[1], D. Bortot[1], M. Huault[2,3], G. Zeraouli[2,3], V. Ospina[2,3], S. Malko[2,3], J. I. Apiñaniz[2], J. A. Perez-Hernández[2], D. De Luis[2], G. Gatti[2], L. Volpe[2,4], A. Pola[1] and M. Passoni*[1]

[1] *Politecnico di Milano, Via Ponzio 34/3, I-20133 Milan, Italy*

[2] *Centro de Laseres Pulsados (CLPU), Edicio M5. Parque Científico. C/ Adaja, 8. 37185 Villamayor, Salamanca, Spain*

[3] *Universidad de Salamanca, Patio de Escuelas 1, 37008 Salamanca, Spain*

[4] *Laser-Plasma Chair at the University of Salamanca, Salamanca, Spain*

*E-mail: francesco.mirani@polimi.it, matteo.passoni@polimi.it

## Abstract

**Among the existing elemental characterization techniques, Particle Induced X-ray Emission (PIXE) and Energy Dispersive X-ray (EDX) spectroscopy are two of the most widely used in different scientific and technological fields. Here we present the first quantitative laser-driven PIXE and laser-driven EDX experimental investigation performed at the Centro de Láseres Pulsados in Salamanca. Thanks to their potential for compactness and portability, laser-driven particle sources are very appealing for materials science applications, especially for materials analysis techniques. We demonstrate the possibility to exploit the X-ray signal produced by the co-irradiation with both electrons and protons to identify the elements in the sample. We show that, using the proton beam only, we can successfully obtain quantitative information about the sample structure through laser-driven PIXE analysis. These results pave the way towards the development of a compact and multi-functional apparatus for the elemental analysis of materials based on a laser-driven particle source.**

Analytical techniques of X-ray emission spectroscopy play a crucial role in many fields of materials science. They rely on the irradiation of samples with ionizing radiation and the detection of the emitted characteristic X-rays. The elements are recognized according to the X-ray energies, while their concentrations are retrieved from the number of counts. The analytical capabilities of a specific technique depend on the type of incident particles. A widespread technique which exploits keV energy electrons is Energy Dispersive X-ray (EDX) (*1*) spectroscopy. EDX is a fast method for the multielemental analysis of solid homogeneous samples. Conventional EDX is performed only in vacuum and the probed thickness is of micrometers. Another powerful non-destructive technique is Particle Induced X-ray Emission (PIXE) (*2, 3*). Exploiting MeV energy protons, PIXE provides both elemental concentrations of homogeneous samples and stratigraphic structures of complex artifacts down to few 10s µm. Unlike EDX, PIXE can be performed also in-air. Since its birth, the extensive use of PIXE has been limited by the use of large accelerators. Both EDX and PIXE are widely exploited in semiconductor industry (*4, 5*), environment monitoring (*6, 7*) and cultural heritage preservation (*8, 9*). These fields could substantially benefit from the adoption of a flexible, multi-particle tool for X-ray emission spectroscopy with different capabilities.

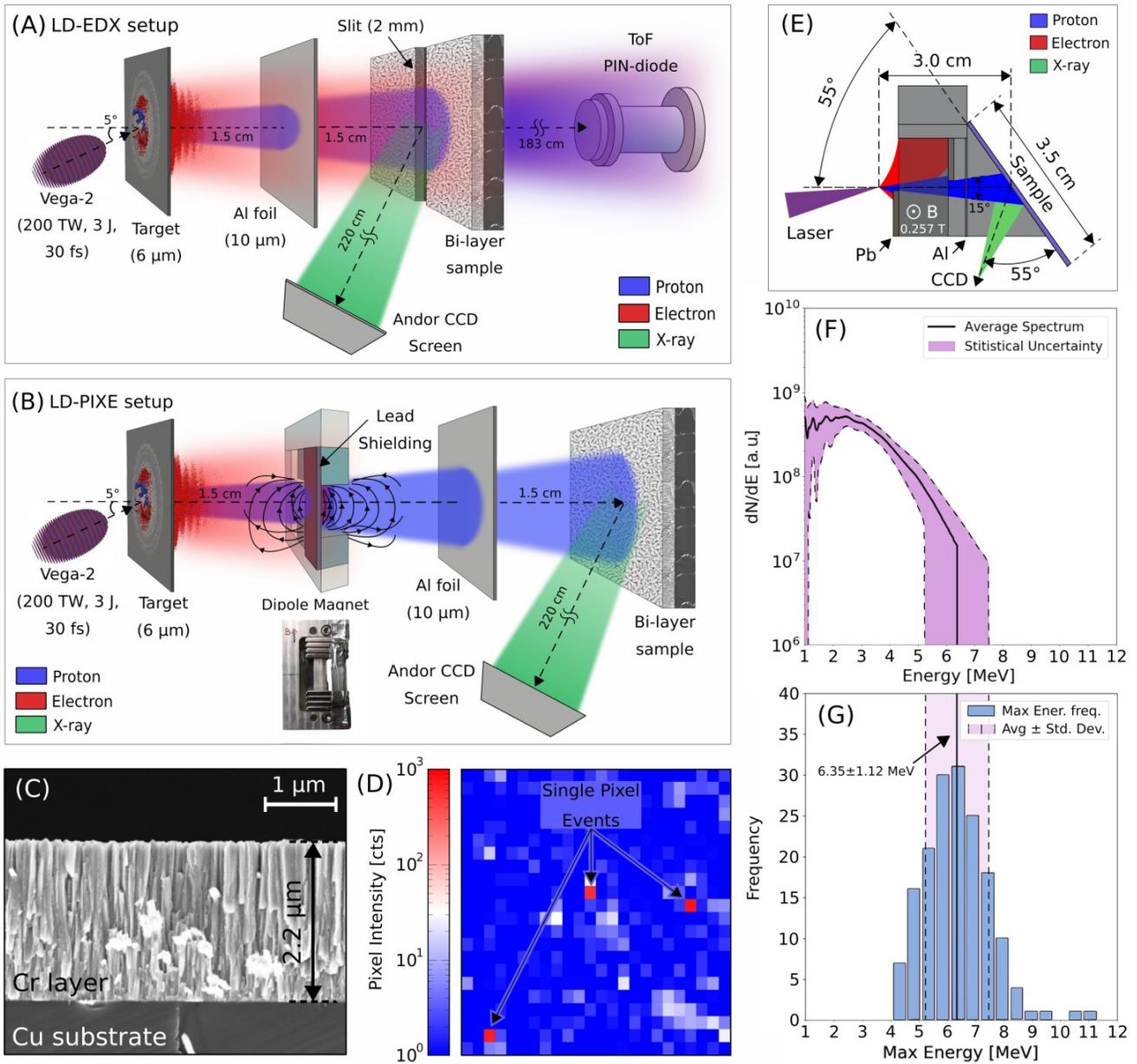

**Fig. 1 Conceptual schemes of the experimental setups. (A)** Schematic illustration for the LD-EDX setup. **(B)** Schematic illustration for the LD-PIXE setup. **(C)** Scanning Electron Microscope (SEM) cross section view of the irradiated sample. **(D)** Detail of a recorded CCD image for X-ray detection. Single pixel events are indicated. **(E)** Top view scheme of the LD-PIXE setup. **(F)** Proton energy spectrum recorded with Time-of-Flight (ToF) spectrometer. No absolute calibration is provided on the vertical axis. The continuous black line is the average spectrum. The purple area represents the statistical uncertainty (i.e. ± standard deviation), provided as the superposition of two separated contributions: the uncertainty on the signal for any energy value and the uncertainty on the maximum proton energy. **(G)** Absolute frequency distribution for the maximum proton energy recorded with ToF measurements. The vertical line represent the mean value, while the purple band width is two times the standard deviation.

Laser-driven sources (*10, 11*) are worth of consideration for materials science applications (*12–15*). A common acceleration scheme exploits the interaction between ultra-short (tens fs) super-intense ($I>10^{18}$ W/cm$^2$) laser pulses and micrometric solid targets to accelerate electrons and protons to energies ranging from few MeVs up to 100s MeV. Electrons and protons are accelerated together in an ultra-fast dynamics and their energy spectra are broad. Because of their peculiar features, compact laser-driven accelerators could be exploited for EDX (*16*) and PIXE (*17–19*). Remarkably, since the range of MeVs energy electrons in solids is of several mm, they could be employed to detect elements deeper inside samples compared to keV electrons. Moreover, MeV electrons can propagate for 100s cm in air, thus enabling ex-situ EDX on large surfaces. Furthermore, the current laser technology provides table-top 10s TW class lasers (*20*) which can accelerate protons up to the energies required by PIXE.

In this work we experimentally perform elemental analysis of a non-homogeneous sample exploiting a laser-driven source. To that end, we present the first *laser-driven EDX (LD-EDX)* and quantitative *laser-driven PIXE (LD-PIXE)* analysis. The experiment was performed at the Centro de Láseres Pulsados with the Vega-2 laser (*21*).

We propose two setups to exploit at best the laser-driven charged particle source, performing the sample irradiation either with both electrons and protons (Fig. 1(A)) or only with protons (Fig. 1(B)). We show that the X-ray yield induced by electron irradiation is dominant in the first configuration (hence we call it *LD-EDX setup*) and it can be exploited to effectively identify the elements. Under the second irradiation condition (*LD-PIXE setup*), we practically demonstrate that a LD-PIXE signal can be used to retrieve quantitative stratigraphic information about the sample structure. We support our experimental results through a Monte Carlo investigation of both setups.

## Results and Discussion

### Laser-driven EDX and laser-driven PIXE experimental setups

In both LD-PIXE and LD-EDX setups, the 200 TW Vega-2 laser pulse interacts with a micrometric aluminium foil to accelerate electrons and protons toward the sample placed in the vacuum chamber. We interpose a second aluminum sheet between the laser-driven particle source and the sample in order to stop the debris produced by the laser-target interaction. The sample has the same composition in both setups. It is made of a 2.2 μm thick layer of chromium deposited onto a 1 mm thick substrate of pure copper. Oxygen is present as a contaminant ($<$ 10 %) in the Cr layer. The film density is equal to 5.3 g/cm$^3$. A cross section view of the sample is shown in Fig. 1(C). Details about the sample production are provided in the Methods section. To detect the emitted X-rays, we exploit a CCD. The experimental setups are designed to allow a post-processing, single photon counting spectra reconstruction (*22, 23*). Single photon events can be clearly distinguished in Fig. 1(D). The CCD energy calibration (see the Methods section) was done through the irradiation of a pure Cu sample with the LD-EDX setup.

Since the LD-EDX setup has the dual purpose of irradiating the sample and characterizing the accelerated protons, we created an aperture slit in the middle of the sample so that a fraction of the protons could reach the ion diagnostics. In the LD-PIXE setup the sample is not splitted and a 0.26 T dipole magnet and lead shields are placed behind the target to remove the electrons (see Fig. 1(E)). Further details about the setups are provided in the Methods section.

The proton diagnostics in the LD-EDX setup is a Time-of-Flight spectrometer (*24, 25*) (see the Methods section), which is aligned with the sample slit along the target normal direction. The energy spectrum averaged over 166 shots, as well as the statistical uncertainty, is shown in Fig. 1(F). In Fig. 1(G) we report the absolute frequency distribution of maximum proton energy (average value 6.35 MeV, standard deviation 1.12 MeV). It is worth mentioning that the knowledge of the shape and the cut-off energy of the proton spectrum is fundamental for LD-PIXE analysis as shown below.

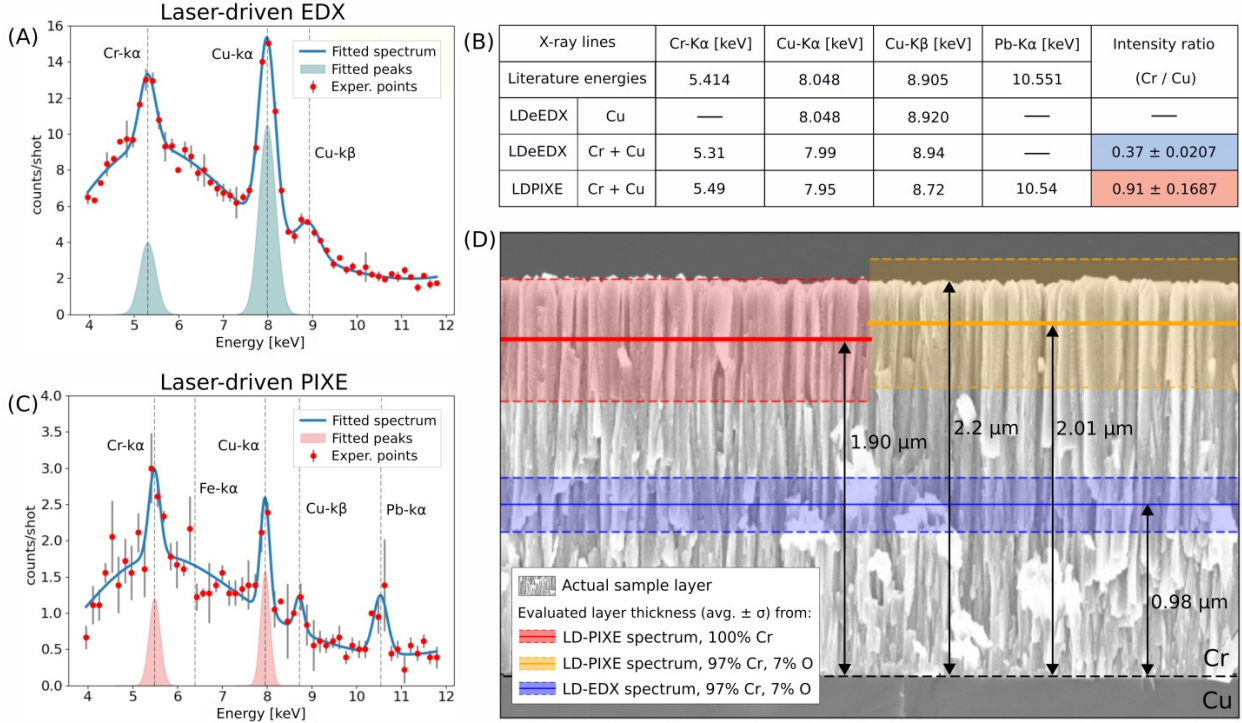

**Fig. 2 Laser-driven electrons EDX / PIXE results. (A)** Recorded spectrum with the LD-EDX setup. The red points are the average over the shots of X-ray intensity at each photon energy, while the length of the error bars is two times the standard deviation. The blue line is the fitted spectrum. The filled blue curves are the Gaussian fits for the peaks. **(B)** Summary of the recorded X-ray line positions and intensity ratios of the Cr and Cu peaks. The first row reports the expected X-ray energies from literature. The second row is related to the X-ray spectrum used to calibrate the CCD (see the Methods section), while the last two rows list the peak positions in the spectra, expressed as the centroid of the fitting Gaussian. **(C)** Recorded spectrum with the LD-PIXE setup. The filled red curves are the Gaussian fits for the peaks. **(D)** Results of the LD-PIXE stratigraphic analysis. The cross section of the film is reported in background. The red, yellow and blue lines are the film thicknesses obtained using the X-ray spectra and film compositions reported in the figure legend.

## Laser-driven EDX elemental and laser-driven PIXE stratigraphic analyses

We start by discussing the LD-EDX spectrum. The goal is to perform *elemental analysis*, i.e. the identification of the elements present in the sample. Then, we focus on the quantitative characterization of the sample structure through LD-PIXE analysis. We call this procedure *stratigraphic analysis*, aimed at determining the Cr layer thickness.

We irradiate the sample with 42 particle bunches (i.e. shots) in the LD-EDX setup. The corresponding X-ray spectrum per unit of shots is presented in Fig. 2(A). As expected, well defined peaks emerge from the background. The fit is performed with the Levenberg-Marquardt least-square fitting algorithm (*26*), a standard method for X-ray spectra interpolation. The peaks are fitted with a Gaussian shape (*27*), while the background is modeled with an exponential polynomial of third order (*27*). Fig. 2(A) shows also the fitted Cr and Cu peaks after background subtraction. In the case of Cr, the Kα and Kβ lines are too close to be distinguished. For the present study the intent is to identify both Cr and Cu elements (the detection efficiency of the shielded CCD is too low at the energy corresponding to the oxygen line). Fig. 2(B) lists the energies of the X-ray peaks. They are very close to the actual values, with a relative deviation always less than 2%. Since we can uniquely identify the elements, we can conclude that LD-EDX provides a reliable elemental analysis.

Unlike the elemental analysis, the stratigraphic analysis requires a model that relates the X-ray yields, the material composition and the energy of the incident particles. As far as LD-EDX is concerned, the electron spectrum characterization is not trivial. Moreover, no analytical models to describe the X-ray emission induced by high energy electrons are available. Therefore, the X-ray spectrum obtained by LD-EDX can not be directly interpreted to retrieve stratigraphic information. To obtain the latter, both reference samples and prior knowledge of the accelerated electron spectrum would be needed (28).

To overcome the aforementioned limitations, we switch to LD-PIXE. The LD-PIXE spectrum (Fig. 2(C)) is obtained with 16 particle bunches. Also in this case several peaks are visible. Besides the Cr and Cu characteristic peaks, the spectrum shows also the Pb-K$\alpha$ signal due to the lead shields. In addition, we can recognize a weak signal at 6.3 keV, likely to be related to the iron in the magnet. As for LD-EDX, all elements are correctly recognized as well. However, the X-ray signal per shot obtained with the LD-PIXE setup is approximately 10 times less intense than that recorded using the LD-EDX setup coherently with published theoretical results (16). Therefore, we conclude that LD-EDX provides the sample elemental analysis with a lower number of shots compared to LD-PIXE.

On the other hand, the X-ray yields obtained with the LD-PIXE setup can be exploited for stratigraphic analysis. We have proposed an analytical model and a procedure to determine the sample structure from a LD-PIXE measurement (17). For the specific case considered here, the mass thickness of the Cr layer can be evaluated solving equation (1) presented in the Methods section. It relates the X-ray intensity ratio (i.e. $Cr/Cu$ reported in Fig. 2(B)), the film composition and the incident proton spectrum shape. Fig. 2(D) shows the comparison between the sample cross section and the reconstructed thickness from LD-PIXE analysis. Assuming a pure Cr film and Cu substrate, i.e. neglecting the presence of oxygen, we estimate a layer thickness of 1.90±0.39 µm using the ratio of Cr and Cu X-ray yields obtained experimentally. Considering also the presence of 7% oxygen in the Cr film, we find a thickness equal to 2.01±0.39 µm, which is even closer to the actual value of 2.2 µm. We evaluated the error through a Monte Carlo approach (29) taking into account the uncertainty of both X-ray and proton spectra (see the Methods section). They contribute almost equally to the overall uncertainty, which can therefore be reduced by increasing the number of shots, optimizing the detection system or improving the proton source reproducibility. The last point can be achieved by optimizing the target manufacturing and the laser stability. Anyhow, even under the present experimental conditions, LD-PIXE provides a satisfactory estimation of the actual thickness. Finally, Fig. 2(D) reports also the thickness evaluated considering the yields ratio obtained with the LD-EDX setup (i.e. assuming to ignore the presence of the incident electrons in the LD-PIXE measurement). The result of 0.98±0.16 µm strongly underestimates the actual chromium thickness. Since electrons are more penetrating compared to protons, they unbalance the yields ratio in favor of the copper.

In light of the obtained results, we can conclude that: i) Because of the high intensity of the X-ray signal, the recommended setup to perform elements identification is the LD-EDX one. ii) On the other hand, the quantitative stratigraphic analysis requires to remove the electrons contribution to the X-ray signal. Therefore, LD-PIXE must be exploited. iii) The knowledge of the absolute number of incident protons is not required to perform LD-PIXE analysis. This is very convenient from the experimental point of view, since no absolute calibration of the proton detector is needed. iv) With a suitable theoretical description of LD-PIXE (17), the monochromaticity of the incident proton spectrum is not required.

We can also draw further fundamental prospects regarding the potential of the approach we are presenting: i) the large range of MeV electrons in solids can open to the possibility of the analysis down to 100s µm depth. At these energies, the electron impact ionization cross sections (30) for the K$\alpha$ shells of heavy elements become significant. Moreover, the associated X-ray energies are of the order of 10s keV, thus subject to weak attenuation in thick layers. These considerations suggest that LD-EDX should allow to recognize the presence of heavy elements in matrices within the mm thickness range, significantly extending the capabilities of EDX. ii) It should be noted that the results here presented for a 200 TW laser should be obtained exploiting compact 10s TW class lasers (31, 32). Indeed, they can provide protons with maximum energy of ~6 MeV. Moreover, compact lasers operate with a repetition rate equal to 10 Hz. Assuming a number of accelerated protons of $10^9$-$10^{10}$ particles per shot and the aforementioned repetition rate, the resulting current is approximately 0.1-1 nA. These values are compatible with the currents exploited in conventional PIXE analysis for cultural heritage studies (*33*).

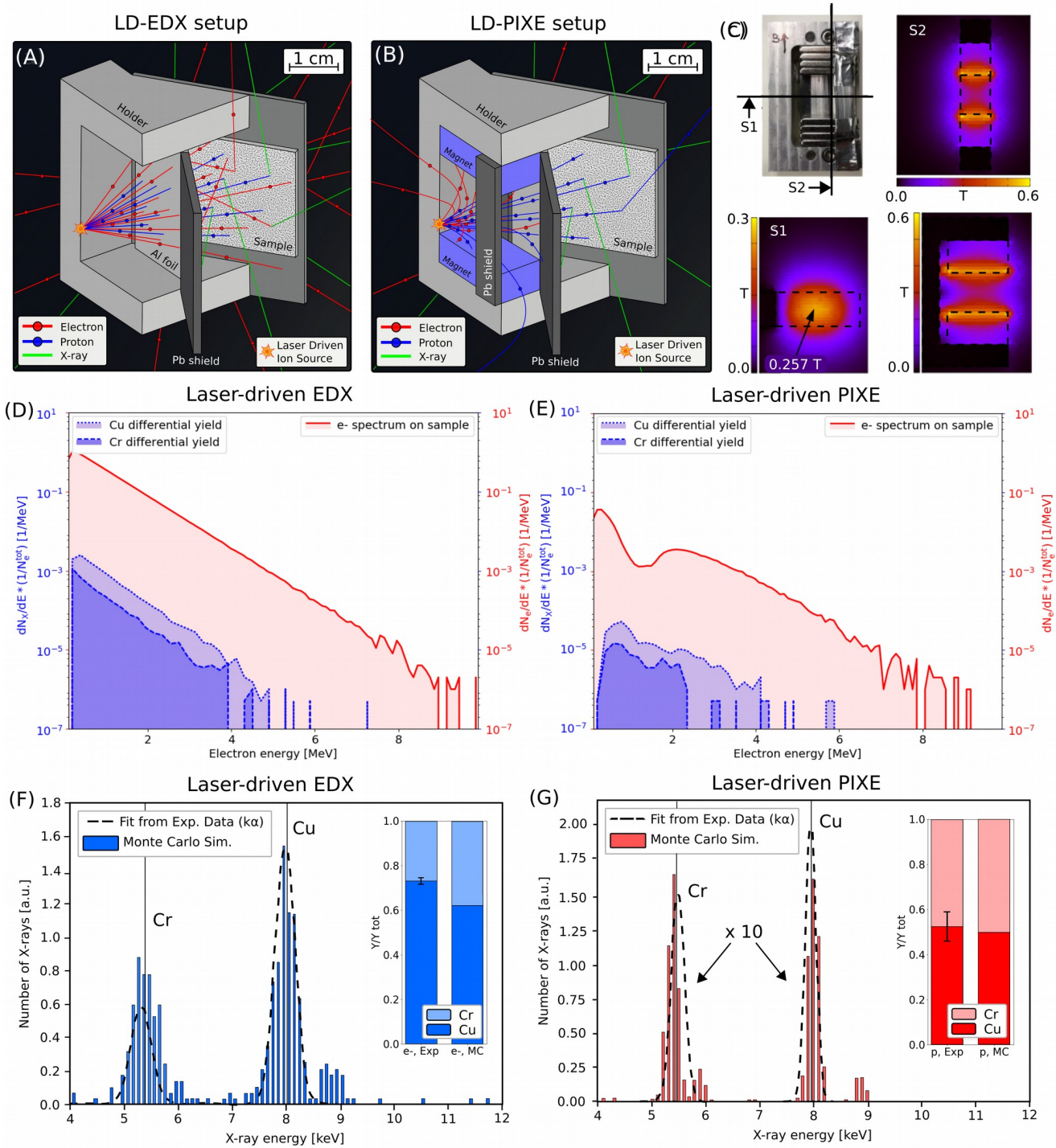

**Fig. 3 FEM and Monte Carlo simulations of laser-driven EDX / PIXE. (A - B)** Snapshots of the LD-EDX and LD-PIXE Monte Carlo simulations. **(C)** Magnetic field intensity map obtained with FEM analysis of the magnet in the LD-PIXE setup. **(D - E)** LD-EDX and LD-PIXE Monte Carlo simulation outputs for electrons as primary particles. The filled red areas are the energy spectra of the electrons incident on the sample surface. The filled blue curves are the X-ray differential yields. **(F - G)** Comparison between the simulated peaks and fits from experimental data for LD-EDX and LD-PIXE. The inset graphs compare the experimental and simulate X-ray yield ratios for LD-EDX and LD-PIXE.

**Assessment of the electrons influence on laser-driven PIXE via Monte Carlo simulations**

In light of the results presented in the previous section, a proper characterization of the system exploited to remove the electrons in the LD-PIXE setup is necessary. Indeed, the thickness obtained from LD-PIXE analysis (i.e. 2.01 µm) underestimates the actual thickness (i.e. 2.2 µm) of about 10%. We can suppose that even in presence of the magnet, in the LD-PIXE setup, a residual fraction of electrons can reach the sample. Therefore, the fraction of X-rays due to the electrons irradiation compared to the amount induced by protons must be properly estimated. Thus, we performed Geant4 (*34*) Monte Carlo simulations of the charged particles propagation, their interaction with the sample and X-ray generation.

We simulate LD-EDX and LD-PIXE analysis considering both protons and electrons as primary particles. The simulated setups are shown in Fig. 3(A, B). Simulations are performed with the same number ($10^8$) of electrons and protons. The primary particles are generated with a uniform angular distribution between ±20°, so that the sample surface is entirely hit by the particles. The proton energies are extracted from the measured energy distribution (Fig. 1(F)). On the other hand, the electron energy spectrum is modeled as an exponential distribution with maximum energy of 10 MeV. We considered two electron temperatures equal to 0.67 MeV and 1.0 MeV. The first value is estimated from the actual laser parameters using a generalized ponderomotive scaling (*35*). The second temperature is obtained by matching the theoretical estimation of the maximum proton energy (provided by the Passoni-Lontano model (*36*)) and the corresponding experimental value. For a detailed description of the equations for the temperature evaluations, please refer to the Methods section. To properly simulate the LD-PIXE setup, we first evaluate the 3D magnetic field intensity distribution by means of a magnetostatic Finite Elements Analysis (*37*) (mFEA). The resulting intensity map is shown in Fig. 3(C). Then, we included the 3D magnetic field distribution in the Geant4 simulation. The mFEA simulation is extensively described in the Methods section and a full presentation of the Monte Carlo code implementation is provided in already published works (*16, 17*).

We start considering the simulations with electrons as primary particles. For the 0.67 MeV electron temperature, the simulated energy spectra for the LD-EDX and LD-PIXE setups are shown in Fig. 3(D, E), respectively. All data are normalized to the total number of simulated primary electrons. Since in the LD-EDX setup particles are not deflected, the spectrum of the electrons impinging on the sample practically coincides with the input exponential one. On the other hand, in LD-PIXE setup the number of electrons reaching the sample surface is drastically lowered due to the effect of the magnet. Overall, about 98% of the incident electrons are removed in the case of 0.67 MeV temperature. With a temperature of 1.0 MeV, 96.3% of the incident electrons are removed. Fig. 3(D, E) show also the X-ray differential yields. They are defined as the number of characteristic X-rays (i.e. at 5.41 keV and 8.05 keV for Cr and Cu, respectively) leaving the sample as a function of the incident electron energy. The differential yields confirm that MeV electrons contribute to the X-ray production because the substrate is thick enough to let them slow down. The ratios between the X-ray yields due to the electrons in presence and in absence of the magnet (i.e. $Y_{LD-PIXE,e}/Y_{LD-EDX,e}$) are 0.020 and 0.036 for the 0.67 MeV and 1.0 MeV electron temperatures, respectively. On the other hand, the experimental ratio between the X-ray yields in LD-EDX and LD-PIXE $((Y_{LD-EDX,e}+Y_{LD-EDX,p})/(Y_{LD-PIXE,e}+Y_{LD-PIXE,p}))$ is equal to 8.6. Neglecting the proton contribution in LD-EDX (i.e. $Y_{LD-EDX,p}$), we can provide a conservative estimation of the residual X-ray yield due to the electron irradiation in the LD-PIXE experiment (i.e. $Y_{LD-PIXE,e}/(Y_{LD-PIXE,e}+Y_{LD-PIXE,p})$). We find that <17% and <30% of the X-rays in the LD-PIXE setup are related to surviving electrons instead of protons for the 0.67 MeV and 1.0 MeV electron temperatures, respectively. This is coherent with a 10% underestimation of the thickness as found in the previous section.

In Fig. 3(F, G) we compare the fit of the experimental X-ray Kα peaks and the Monte Carlo simulated spectra for the LD-EDX (with 0.67 MeV electron temperature) and LD-PIXE spectra, respectively. In the LD-EDX simulation we neglect the proton contribution, while in the LD-PIXE simulation we neglect the electron one. The counts in each channel are normalized with respect to the total number of characteristic X-rays. For both spectra, the experimental and simulated peaks are in good agreement. We can directly compare also the simulated and experimental ratios between chromium and copper yields. They are reported as bar plots in Fig. 3(F, G). Again, we have a reasonable agreement between the experimental data and Monte Carlo simulations. These results give a further evidence of the fact that LD-EDX is dominated by the electron contribution to the X-ray production, while in LD-PIXE the protons are playing the crucial role.

Finally, starting from the experimental and simulated yield ratios, we can provide a post-hoc estimation of the electron temperature. By means of analytical calculation (see the Methods section), we find that this value is 0.63±0.49 MeV. The result practically coincides with the 0.67 MeV electron temperature, we estimated from laser parameters, while the 1.0 MeV estimation falls within its uncertainty.

## Conclusions

We show that a laser-driven particle source can be exploited as a powerful tool to characterize samples of unknown elemental composition. Since this unconventional acceleration scheme provides both high energy electrons and protons, the experimental setup can be adjusted to perform both LD-EDX and LD-PIXE analyses. We show that LD-EDX can be used to successfully identify the elements with fewer shots compared to LD-PIXE. Remarkably, LD-EDX prospects the possibility, not achievable with traditional EDX, to analyze large artifacts in-air and to probe the presence of heavy elements at millimeter depths. Besides, we experimentally demonstrate that LD-PIXE can be used to perform quantitative stratigraphic analysis of a non-homogeneous sample. Thus, LD-EDX and LD-PIXE prove to be complementary techniques for elemental characterization of a sample. Finally, based on our results we suggest that the analyses carried out with a 200 TW laser might be performed also with compact 10s TW class lasers and advanced targetry solutions. Our results represent a significant step towards the development of a compact and versatile laser-based radiation source for multiple materials science investigations.

## Materials and Methods

### Chromium film deposition via Magnetron Sputtering

The Cr film was grown on a pure Cu substrate exploiting an High Power Impulse Magnetron Sputtering Deposition System (38) located in Politecnico di Milano. The deposition was performed in Direct Current (DC) mode. This technique allows to obtain planar films on large surface areas (several cm$^2$). In order to avoid the delamination of the film due to the strong stresses (39) induced by a long deposition time, we broke the deposition process in several steps. For a complete list of the deposition parameters, see the Supplementary materials.

### CCD energy calibration

CCDs are standard diagnostics for spectroscopy in laser-plasma interaction experiments (22, 40, 41). The Andor Ikon-M D0934P-BN CCD camera (1024 × 1024 pixels) energy calibration is performed exploiting the LD-EDX setup shown in Fig, 1(A) and a mono-elemental sample of pure copper. Fig. 4 shows the X-ray spectrum obtained with 16 shots. In order to perform the energy calibration, we first evaluate the local background distribution around any single event pixel. By local background we mean the average intensity related to the eight pixels surrounding a single event pixel. The distribution is reported in the inset graph of Fig. 4. Its shape is Gaussian and it is located around 0 – 10 cts. Since the X-ray peaks of interest lies around 100s cts, we neglect the local background contribution and we directly calibrate the CCD considering the absolute Cu Kα peak position. The final calibration factor is 8.048 keV / 277 cts = 0.029 keV/cts. Using this calibration, we obtain the energies reported in Fig. 2(B) for all spectra presented in this work.

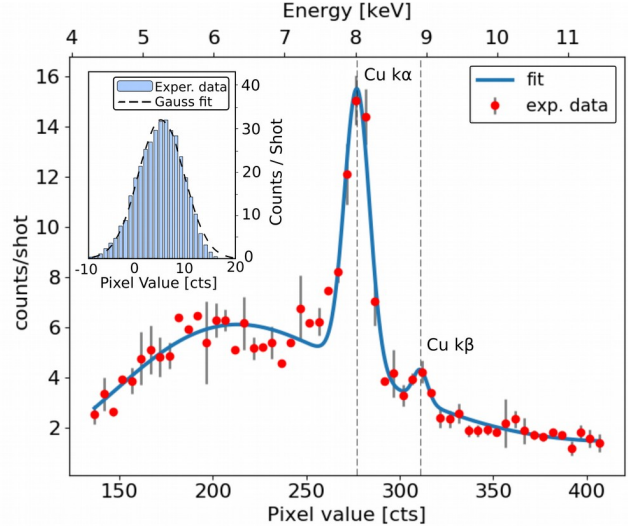

**Fig. 4 Laser-driven EDX spectrum for the CCD energy calibration.** Copper spectrum obtained with the LD-EDX setup used to calibrate the X-ray CCD. The inset graph is the pixel value distribution of the local background around the single pixel events.

### Additional details about the experimental setup

The Vega-2 laser pulse has 30 fs pulse duration and P-polarization. The energy is 3 J on target and the intensity is ~$2\times10^{20}$ W/cm². The spot size FWHM is 7.0 μm. The angle of incidence with respect to the target normal is 5°. The thickness of the aluminium target and the protective sheet are 6.0 μm and 10.0 μm, respectively. As far as the sample is concerned, the chromium film was grown directly on the copper substrate via Magnetron Sputtering (see previous sub-section). The actual composition is 93% of chromium and 7% oxygen (expressed as mass fraction). The reported film density equal to 5.3 g/cm³ has been measured by means of conventional EDX (28) and weight difference between the sample and bare substrate. In the LD-EDX setup the sample slit is 2.2 μm thick. We place a lead shield to protect the CCD screen from the radiation emitted at the laser-target interaction point. The screen is further protected with 2 μm thick Mylar and 6 μm thick aluminium foils.

### Time-of-Flight spectrometer

A Time-of-Flight (ToF) spectrometer has been used to characterize the proton spectra. Its location in the experimental setup is shown in Fig. 1(A). As ToF diagnostics, we exploited a 1 ns time-resolved pin diode detector. To perform the analysis of the recorded signal, we adopt the approach provided by *Milluzzo et al.* (24).

### Model for Laser-driven PIXE quantitative analysis

For the specific case considered in this work, the mass thickness of the Cr layer $(\varrho t)_{Cr}$ can be evaluated solving the following equation.

$$\frac{Y_{Cr}}{Y_{Cu}} = \frac{\varepsilon_{Cr}}{\varepsilon_{Cu}} \frac{M_{Cu}}{M_{Cr}} \frac{W_{Cr}}{W_{Cu}} \times \frac{\int_{E_{p,max}}^{E_{p,min}} f_p(E_p) \int_{E_{p,out\,Cr}}^{E_p} \sigma_{Cr}(E)\omega_{Cr} e^{-\mu_{Cr}\int_E^{E_p}\frac{dE'}{S_{Cr}(E')}\frac{\cos\theta}{\cos\varphi}} \frac{dE}{S_{Cr}(E)} dE_p}{e^{-\left(\frac{\mu}{\rho}\right)_{Cr}\frac{(\rho t)_{Cr}}{\cos\varphi}} \int_{E_{p,max}}^{E_{p,min}} f_p(E_p) \int_{E_0}^{E_{p,out\,Cr}} \sigma_{Cu}(E)\omega_{Cu} e^{-\mu_{Cu}\int_E^{E_p}\frac{dE'}{S_{Cu}(E')}\frac{\cos\theta}{\cos\varphi}} \frac{dE}{S_{Cu}(E)} dE_p} \quad (1)$$

where $Y_{Cr}/Y_{Cu}$ is the ratio of Cr and Cu X-ray yields obtained experimentally (blue cell in Fig. 2(B)); $\varepsilon_{Cr}/\varepsilon_{Cu}$ is the ratio of CCD efficiencies at the Cr and Cu Kα energies. $\varepsilon_{Cr}$ and $\varepsilon_{Cu}$ account for both the CCD quantum efficiency (from the instrument documentation) and the attenuation due to the Mylar and aluminium foils. $M_{Cu}/M_{Cr}$ is the atomic mass ratio; $W_{Cu}/W_{Cr}$ is the ratio of the Cu and Cr mass concentrations in the film and substrate respectively; $E_{p,min}$ and $E_{p,max}$ are the minimum and maximum incident proton energy; $f_p(E_p)$ is the proton spectrum shown in figure 1(f); $E_{p,out\ Cr}$ is the proton energy at the interface between Cr layer and Cu substrate; $\sigma_{Cr}$ and $\sigma_{Cu}$ are the ionization cross sections (42, 43); $\omega_{Cr}$ and $\omega_{Cu}$ are the fluorescence yields (44); $\mu_{Cr}$ and $\mu_{Cu}$ are the X-ray attenuation coefficients (evaluated with XCOM (45) code); $\vartheta$ is the proton incidence angle; $\varphi$ is the X-ray emission angle; $S_{Cr}$ and $S_{Cu}$ are the proton stopping power in the Cr layer and Cu substrate (from SRIM (46) code). The proton mass range inside the sample is linked to the energy with the stopping power trough the relation $S(E_p)=-dE/d(\varrho t)$.

Since both $Y_{Cr}/Y_{Cu}$ and $f_p$ are known, equation (1) can be solved numerically to find the mass thickness of the Cr layer $(\varrho t)_{Cr}$. Because the problem is strongly non-linear, the solution must be found iteratively. For a more general description of both the analytical model and the iterative algorithm, see ref. (17).

**Error evaluation**

In order to combine the uncertainties on the incident proton spectrum $fp(Ep)$ and on the yields ratio $Y_{Cr}/Y_{Cu}$, the measurement error on the layer thickness is evaluated with a Monte Carlo approach. Briefly, we calculate several times the thickness of the sample by means of equation (1) and the procedure presented in ref. (17), changing $fp(Ep)$ and $Y_{Cr}/Y_{Cu}$ at each evaluation. $fp(Ep)$ are extracted from the set of recorded spectra. $Y_{Cr}/Y_{Cu}$ values are extracted form a Gaussian distribution. The mean values are 0.91 and 0.37 and the standard deviations are 0.1687 and 0.0207 for LD-PIXE and LD-EDX, respectively. These values are also reported in Fig. 2(B). To obtain the thickness distribution for the LD-PIXE shown in Fig. 5(A), we evaluated the thickness $10^3$ times. Coherently with the Central Limit Theorem, the resulting thickness distribution tends to a Gaussian function. We consider its standard deviation as the thickness error. Fig. 5(B) reports the scatter plot of the data obtained with this procedure. This representation is useful to compare the $fp(Ep)$ and $Y_{Cr}/Y_{Cu}$ contributions to the total uncertainty. Some data obtained in correspondence of the average proton spectrum (red points) and at fixed yields ratio (green points) are superimposed to the heat map. Since the range covered by red and green points along the vertical direction are comparable, we can conclude that the uncertainties contribute evenly to the total error.

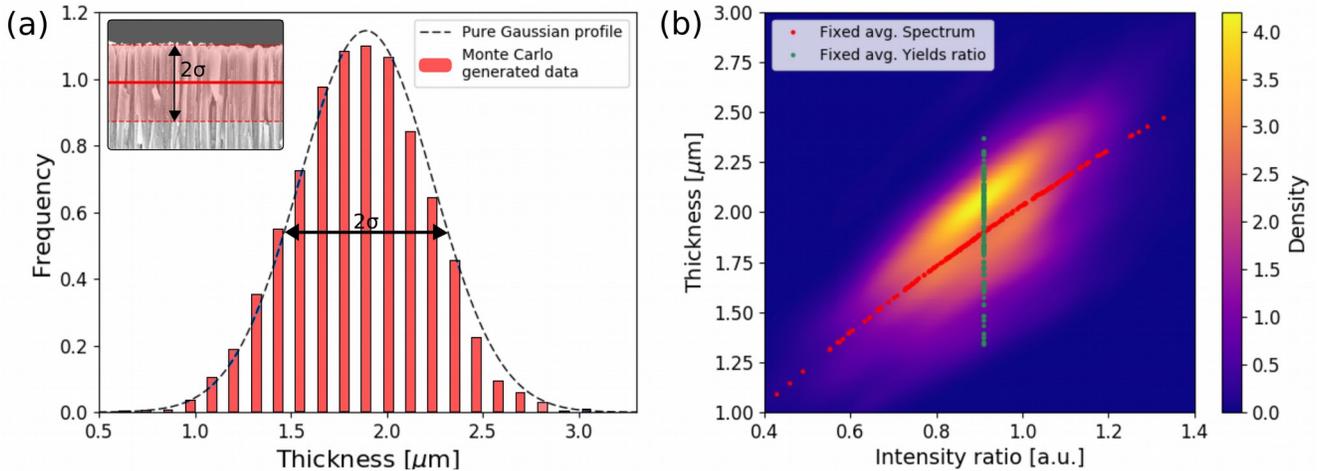

**Fig. 5 Uncertainty evaluation for the laser-driven PIXE measurement.** (A) Frequency distribution of the thickness resulting from the Monte Carlo simulation for the error calculation. The inset image shows the related uncertainty for the thickness measurement. (B) Heatmap scattered data. The extracted intensity ratio is reported on the x-axis, while the resulting thickness is on the y-axis. The color scale is related to the number of occurrences. The red points are obtained by fixing the proton spectrum as the average one. The green points are obtained by fixing the intensity ratio as the average value.

**Theoretical evaluations of the electron temperature for Monte Carlo simulations**
The electron temperatures of 0.67 MeV and 1.0 MeV exploited in the Monte Carlo simulations are evaluated considering two different approaches.
The first value of 0.67 MeV is calculated from the actual laser parameters and the following extended ponderomotive scaling (35).

$$T_e = C_1(a_0, pol, t_{foil}) \cdot 0.511 \cdot \left[\sqrt{1 + \frac{a_0^2}{2}} - 1\right] + C_2(a_0, pol, t_{foil}) \cdot 0.511 \cdot \left[\sqrt{(1 + f^2 \frac{a_0^2}{2})\sin^3\theta} - 1\right]\tan\theta \quad (2)$$

Where $C_1$ and $C_2$ are weighting factors equal to 0.22 and 0.04 respectively, $a_0 = 0.85\sqrt{I\lambda^2/(10^{18}\,W\,cm^{-2}\,\mu m^2)} \cong 9.5$ is the normalized laser amplitude, $\vartheta$ is the laser incidence angle, $f = 1 + \sqrt{1-\eta}$ is the reflection amplification factor with $\eta \approx 0.1$ the conversion efficiency of laser energy into hot electrons (which is a reasonable value in our conditions).
The second value of 1.0 MeV is obtained by matching the maximum proton energy measured in our experiment (i.e. $\approx$ 6.35 MeV) with its estimation obtained through the quasi-static model described in ref. (36) via the following approximated formula:

$$E_{p,max} \approx \left[\log\frac{n_{h0}}{\tilde{n}} - 1\right]T_e \quad (3)$$

where $n_{h0} = 5 \times 10^{20}\,cm^{-3}$ is the hot electron density and $\tilde{n} = 3.4 \times 10^{17}\,cm^{-3}$ is a fitting parameter (both computed following, again, the reasoning presented in ref. (36)).

**Magnetostatic Finite Elements Analysis simulation**
We simulate the 3D static magnetic field created by the structure shown in Fig. 3(B) with the finite elements library Sparselizard (37). A similar 2D example (47) is widely present in the Sparselizard libraries. Please refer to the example for a detailed description of the electrostatic problem and its implementation. Our magnets are characterized by a magnetization equal to $8.0 \times 10^5$ A/m. For the permeability of the various regions we assume $4\pi \times 10^{-7}$ H/m in vacuum and magnet volumes, while in the support volume we take $10^3$ times the value in vacuum. The structure is placed at the center of a cubic box (10 cm face). The simulated mesh is generated with the Gmesh program (48) and it is formed by approximately $2 \times 10^5$ nodes. In Fig 3(C) the resulting magnetic field intensity is mapped on three different planes. The magnetic field at the center of the magnet obtained with the simulation coincides with the actual measured value of 0.26 T. Finally, we evaluate the magnetic field intensity at the nodes of a 10x10x10 grid placed between the magnetic plates. In this way we cover the region of interest where the electrons and ions can travel and, eventually, be deflected. The grid is fed as input in the Geant4 Monte Carlo simulations.

**Post-hoc estimation of the electron temperature**
Exploiting Monte Carlo simulations, the ratio between the copper X-ray yields due to electrons in LD-PIXE and LD-EDX can be fitted as a function of the temperature $T_e$ as $Y^{Cu}_{LD-PIXE,e}/Y^{Cr}_{LD-EDX,e} = 0.035 \times T_e + 0.0034$. The linear fit is shown in Fig. 6. From Monte Carlo simulations, the ratio between the chromium and copper X-ray yields due to electrons in LD-PIXE setup is $a = Y^{Cr}_{LD-PIXE,e}/Y^{Cu}_{LD-PIXE,e} = 0.56$. For reasonable values of $T_e$, this ratio remains constant. As far as protons are concerned, the ratio is equal to $b = Y^{Cr}_{LD-PIXE,p}/Y^{Cu}_{LD-PIXE,p} = 1.023$. From the experimental data, the Cr over Cu X-ray intensity ratio obtained with LD-PIXE is

$c = (Y^{Cr}_{LD-PIXE,e} + Y^{Cr}_{LD-PIXE,p})/(Y^{Cu}_{LD-PIXE,e} + Y^{Cu}_{LD-PIXE,p}) = 0.91 \pm 0.169$. Finally, neglecting the X-ray contribution due to protons in the LD-EDX spectra we have $d = (Y^{Cu}_{LD-PIXE,e} + Y^{Cu}_{LD-PIXE,p})/Y^{Cu}_{LD-EDX,e} = 0.083 \pm 0.011$. Combining the presented equations, the electron temperature can be evaluated as

$$T_e = 28.63 \times d \times \frac{b-c}{b-a} - 0.096 = 0.63 \pm 0.49 \, MeV \qquad (4)$$

and the corresponding copper X-ray yields ratio results 0.025±0.017. This point is also reported in Fig. 6.

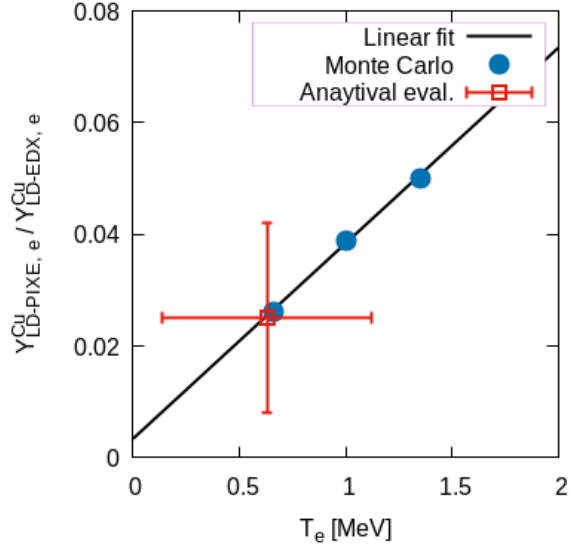

**Fig. 6 Linear fit of the electron temperature as a function of the ratio between the copper X-ray yields due to electrons in laser-driven EDX and laser-driven PIXE.** The blue points are the Monte Carlo simulated data. The black line is the linear fit. The red point is the copper X-ray yields ratio evaluated with the linear fit in correspondence of the electron temperature resulting from the analytical procedure presented in the Methods section.

**Data availability**
The data supporting the findings of this work are available from the corresponding authors on request.

**Author contribution statement**
F.M. designed the experiment, took part to the experiment, performed the analysis of the data, prepared the manuscript. A.M. designed the experiment, took part to the experiment and revised the manuscript. F.C. took part to the experiment, performed the analysis of the data and revised the manuscript. A.P. and A.F. took part to the experiment and revised the manuscript. D.D. designed the experiment, production and characterization of the samples, revised the manuscript. V.R. designed the experiment, provided experimental support and revised the manuscript. D.V. collaborated to the production and characterization of the samples, revised the manuscript. D.B. provided experimental support and took part to the analysis of the data. M.H. prepared the experiment and performed the alignment procedure. G.Z. performed the alignment and X-ray diagnostic. V.O. prepared the Time of Flight. S.M. performed the X-ray Camera calibration. J.I.A. prepared the Thomson Parabola and took part to the experiment. J.A.P. prepared the setup and took part to the experiment. D.D.L. provided technical support. G.G. provided experimental support. L.V. designed the experiment, provided experimental support and revised the manuscript. A.P. provided experimental support, took part to the analysis of the data and revised the manuscript. M.P. conceived the project, took part to the experiment as PI of the campaign at CLPU, supervised all the activities and revised the manuscript.

**Acknowledgments**

This project has received funding from the European Research Council (ERC) under the European Union's Horizon 2020 research and innovation programme (ENSURE grant agreement No 647554). We acknowledge i) the CLPU for granting access to its facilities. ii) The CLPU laser, engineer and administrative teams for all the support in the experiment implementation. iii) The Spanish Ministerio de Economía y Competitividad through the PALMA Grant No. FIS2016-81056-R, the Spanish Ministerio de Ciencia, Innovación y Universidades ICTS Equipment Grant No. EQC2018-005230-P. iv) The LaserLab Europe V Grant No. 871124, and v) the Junta de Castilla y León Grant No. CLP087U16. A special thanks to i) Robert Fedosejevs for lending X ray CCD camera and for the fruitful discussions. ii) Luca Fedeli for supporting the preparation of the proposal for the experiment. iii) Margherita Zavelani for the support to the discussion of the results.

**Competing interests**

The authors declare no competing interests.